\documentclass[useAMS,usenatbib,a4]{mn2e}

\usepackage{times}
\usepackage{graphicx,bm,amssymb}
\usepackage{ulem}
\usepackage{epsfig}
\usepackage{amssymb}
\usepackage{natbib}
\usepackage{psfrag}
\def\araa{ARA\&A}
\def\apj{ApJ}
\def\apjl{ApJ}
\def\apss{Ap\&SS}
\def\aap{A\&A}
\def\mnras{MNRAS}
\def\na{New A}
\def\prd{Phys.~Rev.~D}
\def\pasj{PASJ}
\def\physrep{Phys.~Rep.}

\newcommand{\be}{\begin{equation}}
\newcommand{\ee}{\end{equation}}
\newcommand{\bary}{\begin{eqnarray}}
\newcommand{\eary}{\end{eqnarray}}
\newcommand{\en}{E_\nu}

\def\bi{\begin{itemize}}
\def\ei{\end{itemize}}
\def\lsim{\mathrel{\rlap{\lower3pt\hbox{\hskip1pt$\sim$}}
     \raise1pt\hbox{$<$}}} 
\def\gsim{\mathrel{\rlap{\lower3pt\hbox{\hskip1pt$\sim$}}
     \raise1pt\hbox{$>$}}} 
\def\la{\langle}

\begin{document}
\title[Gamma-ray fluxes from the core emission of  Centaurus A:  A puzzle solved]
{Gamma-ray fluxes from the core emission of  Centaurus A:  A puzzle solved}
\author[N. Fraija]%
{Nissim Fraija \thanks{E-mail:nifraija@astro.unam.mx. Luc Binette-Fundaci\'on UNAM Fellow.}\\
Instituto de Astronom\' ia, Universidad Nacional Aut\'onoma de M\'exico, Circuito Exterior, \\C.U., A. Postal 70-264, 04510 M\'exico D.F., M\'exico}


\maketitle
	
\begin{abstract}
A high-energy component in the radio galaxy Centaurus A was reported after analyzing four years of Fermi data.  The spectrum of this component is described by means of a broken power law with a break energy of 4 GeV and, below and above spectral indices of $\alpha_1$=2.74$\pm$0.03 and $\alpha_2$=2.09$\pm$0.20, respectively. Also a faint $\gamma$-ray flux at TeV energies was detected by H.E.S.S.. In this paper we show that the spectrum at GeV-TeV energies is described through synchrotron self-Compton emission up to a few GeV ($\sim$ 4 GeV) and $\pi^0$ decay products up to TeV energies, although the emission of synchrotron radiation by muons could contribute to the spectrum at GeV energies, if they are rapidly accelerated.  Muons and  $\pi^0$s are generated in the interactions of accelerated protons with two populations of seed photons which were reported by Compton Gamma-Ray Observatory: one population at intermediate state emission with energy peak of 0.15 MeV and another at low state emission with energy peak of 0.59 MeV.  In addition, we show that the reported observations of ultra-high-energy cosmic rays and non high-energy neutrino detection around Centaurus A can be explained through these interactions, assuming that proton spectrum is extended up to ultra-high-energies.
  
\end{abstract}


\section{Introduction}
Centaurus A (Cen A) is classified as Fanaroff $\&$ Riley Class I  \citep{1974MNRAS.167P..31F} active galactic nucleus (AGN). At a distance of D$_z\simeq3.8$ Mpc, it has been  one of the best studied extragalactic sources,  characterized by having an off-axis jet  of viewing angle which is estimated  as $\sim 45^\circ$ \citep[see, e.g.] [and reference therein]{2006PASJ...58..211H} and two giant radio lobes of hundreds of kiloparsec.  Close to the core, this source has been imaged in radio, infrared, optical \citep{1975ApJ...199L.139W,1976ApJ...206L..45M,1970ApJ...161L...1B,1981ApJ...244..429B},  X-ray and $\gamma$-rays (MeV-TeV) \citep{2003ApJ...593..169H,1999APh....11..221S,2009ApJ...695L..40A, 2010ApJ...719.1433A,2011A&A...531A..30R}. Particularly,  in the range of MeV - TeV energies,  Cen A  has been observed by Compton Gamma-Ray Observatory (CGRO) mission, Fermi Gamma-Ray Space Telescope and High Energy Stereoscopic System (H.E.S.S.) experiment.    Observations from OSSE, COMPTEL and EGRET, all of them instruments of Compton Gamma-Ray Observatory (CGRO),  imaged Cen A  from 1991 to 1995 in two states (intermediate and low).  The intermediate state exhibited its brightest peak at $0.15^{+0.03}_{-0.02}$ MeV with a flux of $4.52\times 10^{-4}$ MeV/cm$^2$/s and luminosity of 5$\times 10^{42}$ erg/s whereas the low one showed it at  $0.59^{+0.02}_{-0.02}$ MeV with a flux of $2.91\times 10^{-4}$ MeV/cm$^2$/s and  luminosity of 3$\times 10^{42}$ erg/s \citep{1998A&A...330...97S}.  Also, for a period of 10 months, Cen A was monitored by the Large Area Telescope (LAT) on board the Fermi. The $\gamma$-ray flux collected  was described by a power law with a photon index of 2.67$\pm 0.10_{stat} \pm 0.08_{sys}$ \citep{2010ApJ...719.1433A}.  For more than 120 hr, Cen A was also observed by H.E.S.S. \citep{2005A&A...441..465A, 2009ApJ...695L..40A}.   The spectrum was described by a power law with a spectral index of 2.7$\pm\, 0.5_{stat} \pm 0.2_{sys}$ and an integral flux of $\sim 2.6 \times 10^{39}$ erg s$^{-1}$  above an energy threshold of $\sim$ 250 GeV.    Recently,  \citet{2013ApJ...770L...6S} reported  a new high-energy (HE) component in the spectrum of Cen A as a result of  analyzing four consecutive years of Fermi-LAT data.  The spectrum of this component was described using  a broken power law  with a break energy at  4 GeV and spectral indices  of $\alpha_1$=2.74$\pm$0.03 (below) and $\alpha_2$=2.09$\pm$0.20 (above). It has to be highlighted that although the spectral energy distribution (SED) has been successfully described up to GeV energy range by means of synchrotron self-Compton (SSC) emission,  there is still controversy about the emission processes that could contribute  to spectrum at GeV-TeV energies such us one-zone SSC model  \citep{2010ApJ...719.1433A},  photo-disintegration of heavy nuclei \citep{2013arXiv1312.7440K} and  photo-hadronic processes \citep{2012ApJ...753...40F, 2012PhRvD..85d3012S, 2014A&A...562A..12P}. On the other hand, Pierre Auger Observatory (PAO)  associated some ultra-high-energy cosmic rays  with the direction of Centaurus A  (UHECRs)   \citep{2007Sci...318..938P, 2008APh....29..188P} and IceCube reported 28 neutrino-induced events in a TeV - PeV energy range,  although none of them related with this direction \citep{2013arXiv1311.5238I, 2013arXiv1304.5356I}.\\
In this work, we describe a hadronic model  through  the proton-photon (p$\gamma$) interactions to describe the $\gamma$-ray spectrum from the core emission of Cen A at the GeV - TeV energy range.   Based on the p$\gamma$ interactions and  taking into account the seed photons at two emission states (intermediate and low), we present the synchrotron radiation by muons and $\pi^0$ decay products.  We use the muon synchrotron emission to fit the spectrum up to $\sim$ 126 GeV and $\pi^0$ decay products up to a few TeV.  Finally, correlating the produced $\gamma$-ray fluxes for both states with UHECR and neutrino fluxes, we estimate the number of expected events in PAO and IceCube, respectively. We have assumed that the proton spectrum is extended through a simple  power law up to ultra-high-energies. We hereafter use  k=$\hbar$=c=1 in natural units and z$=0.00183 \pm 0.00002\simeq 0$.
\section{P$\gamma$ interactions}
Relativistic protons are accelerated and cooled down  by p$\gamma$ interactions at the emission region in the jet.  The single-pion production channels  are  
 \begin{eqnarray}
p\, \gamma &\longrightarrow&
\Delta^{+}\longrightarrow
\left\{
\begin{array}{lll}
p\,\pi^{0}       &&   \mbox{fraction }2/3, \\
n\,  \pi^{+}     &&   \mbox{fraction }1/3,
\end{array}\right. \\
\end{eqnarray}
where $n$, $\pi^+$ and $\pi^0$ are neutron, charged and neutral pions, respectively.  Two important quantities that define the efficiency of this process are  the photon density and the optical depth which are given by
\be
n^{obs}_{\gamma}\simeq\frac{ L^{obs}_\gamma}{4\pi\,r^2_d\,\epsilon^{obs}_{\gamma,b}}
\label{den}
\ee
and
\be
\tau_{p}\simeq r_d\, n^{obs}_{\gamma} \sigma_{p\gamma}/\delta_D,
\label{opt}
\ee
here  $r_d\simeq\delta_D\, dt^{obs}$ is the comoving dissipation radius which is limited by the variability time scale $dt^{obs}$, $\sigma_{p\gamma}$ is the cross section  for the production of the delta-resonance in proton-photon interactions, $L^{obs}_\gamma$ is the observed luminosity, $\epsilon^{obs}_{\gamma,b}$ is the break energy of observed photons  and $\delta_D=[\Gamma(1-\beta \mu)]^{-1}$ is the Doppler factor with $\theta$ the observing angle along the line of sight and $\beta\simeq$ 1.    Also we define the optical thickness to pair creation as \citep{1997iagn.book.....P}
\be
\tau_{\gamma\gamma}=\left(\frac{L^{obs}_\gamma}{4\pi r_d^2\,m_e}\ \right)\sigma_T\,r_d/\delta_D\,,
\ee
with m$_e$ the electron mass.  Here we have taken into account that the cross-section for pair production reaches a maximum value close to the Thomson cross section $\sigma_T=6.65\times 10^{-25}\,{\rm cm}^2$.
\subsection{Muon synchrotron radiation}
As known from p$\gamma$ interactions, charged pions and then muons, positrons and neutrinos,  $\pi^{+}\rightarrow \mu^+\,\nu_{\mu}\rightarrow e^{+}\,\nu_{e}\,\overline{\nu}_{\mu}\,\nu_{\mu}$ are produced.  Before muons in a magnetic field of the order of Gauss decay, they could be rapidly  accelerated for a short period of time, radiate photons by synchrotron emission and contribute to the flux at GeV energies \citep{1998PhRvD..58l3005R,2000NewA....5..377A,1993A&A...269...67M, 2011ApJ...736..131A}.    It is useful and  convenient to define the observed photon energies radiated by muons as a function of electron energies.  From the comoving photon energy radiated by muons, $\epsilon'_{\gamma}=\frac{3\pi q_e\,B'}{8\,m_{\mu}^3}\,E_\mu^{'2}$,  the relationship between electron and muon Lorentz factors $\gamma_{\mu}=m^2_\mu/m^2_e\,\gamma_{e}$,  the cooling time characteristic for this process, $t'_{syn,\mu}=6\pi m_{\mu}^4/(\sigma_T\,m^2_e\,B'^2\, E'_{\mu})$ and the maximum acceleration time scale,  $t'_{syn,max}=16\,E'_{\mu}/(3\,q_e\,B')$, we can write the break and maximum photon energies as
\bary\label{ene_ph}
\epsilon^{obs}_{\gamma,c} &=& \frac{m^5_\mu}{m^5_e} \epsilon^{obs}_{\gamma,c-e}\cr
\epsilon^{obs}_{\gamma, max} &=& \frac{m_\mu}{m_e}\epsilon^{obs}_{\gamma, max-e}\,.
\eary
Here m$_\mu$ is the muon mass, $q_e$ is the  elementary charge,  $B'$ and $E'_\mu$ are the  magnetic field and muon energy in comoving frame, respectively.    Supposing that  accelerated muons with energies $\gamma_\mu\,m_\mu$ are  well described by a broken power-law $N_\mu(\gamma_\mu)$:  $\gamma_\mu^{-\alpha}$ for $\gamma_\mu < \gamma_{\mu,b}$ and $\gamma_{\mu,b} \gamma_\mu^{-(\alpha+1)}$ for $\gamma_{\mu,b} \leq  \gamma_\mu<\gamma_{\mu,max}$, then the observed synchrotron spectrum can be written as \citep{1994hea2.book.....L,  2001MNRAS.326.1499H,  2011MNRAS.415..133H}
{\small
\bary
\label{espsynm}
&&\left(\epsilon^2\,\frac{dN}{d\epsilon}\right)_{syn,\gamma}= A_{syn,\gamma-\mu}\cr
&&\hspace{0.4cm}\times  \cases {
\left(\frac{\epsilon^{obs}_{\gamma,c}}{\epsilon_0}\right)^{-1/2} \left(\frac{\epsilon_\gamma}{\epsilon_0}\right)^{-(\alpha-3)/2} &  $\epsilon^{obs}_\gamma < \epsilon^{obs}_{\gamma,c}$,\cr
\left(\frac{\epsilon_\gamma}{\epsilon_0}\right)^{-(\alpha-2)/2}           &  $\epsilon^{obs}_{\gamma,c} < \epsilon^{obs}_\gamma < \epsilon^{obs}_{\gamma,max} $\,,\cr
}
\eary
}
with
\be
A_{syn,\gamma-\mu}=\frac{P^{obs}_{\nu,max}\,n_\mu}{4\pi D^2_z}\epsilon^{obs}_{\gamma,c}\,,
\ee
where $n_\mu=N_\mu\times V$ is the total number of radiating muons in the $V=4\pi r_d^3/3$ and $P^{obs}_ {\nu,max} \simeq \frac{dE/dt}{\epsilon_\gamma(\gamma_\mu)}$ is the maximum radiation power.  Hence, the muon density can be written as
\be\label{dmuon}
N_\mu=\frac{12\pi^4 q_e\,m_e^3}{\sigma_{T}\,m^4_\mu}\,r^3_d\,D^2_z\,\delta^{-1}_D\,\epsilon^{obs,-1}_{\gamma,c-e}\,B^{'-1}\,A_{syn,\gamma-\mu}\,,
\ee
The previous equation gives the information on the muon density as a function of magnetic field.
\subsection{$\pi^0$ decay products}
Neutral pion decays into photons, $\pi^0\rightarrow \gamma\gamma$, and typically carries $20\% (\xi_{\pi^0}=0.2)$ of the proton's energy $E_p$.   As has been pointed out by \citet{2012ApJ...753...40F} and \citet{PhysRevLett.78.2292},  its spectrum can be  derived from the spectral characteristics  of the accelerated protons $dN_p/dE_p=A_p E^{-\alpha}_p$ where $A_p$ is the proton proportionally constant and $\alpha$ the spectral index, target photons ($dn_\gamma/d\epsilon_\gamma$), and  the time scales involved in this interaction  (the dynamical ($t'_d\simeq r_d/\delta_D$) and pion ($t'_{\pi^0}$) cooling time scale). The energy loss rate due to pion production (pion cooling time) can be written as
\begin{equation}
t'_{\pi^0}=\frac{1}{2\,\gamma_p}\int\,d\epsilon\,\sigma_\pi(\epsilon)\,\xi_{\pi^0}\,\epsilon\int dx\, x^{-2}\, \frac{dn_\gamma}{d\epsilon_\gamma} (\epsilon_\gamma=x)\,,
\end{equation}
where $\gamma_p$ is the proton Lorentz factor, $\sigma_\pi(\epsilon_\gamma)=\sigma_{peak}\approx 9\times\,10^{-28}$ cm$^2$ is the cross section of pion production in the $\Delta\epsilon_{peak}$=0.2 GeV at $\epsilon_{peak}\simeq$ 0.3 GeV.  Comparing the pion cooling and  the dynamical time scales ($t'_d/t'_{\pi^0}$),
{\small
\bary
&&f_{\pi^0} \simeq \frac{L^{obs}_\gamma\,\sigma_{peak}\,\Delta\epsilon_{peak}\,\xi_{\pi^0}}{8\pi\,\delta_D^3\,r_d\,\epsilon^{obs}_{\gamma,b}\,\epsilon_{peak}} \cr
&&\hspace{1.5cm}\times\cases{
\left(\frac{\epsilon_{\pi^0,\gamma,c}}{\epsilon_{0}}\right)^{-1} \left(\frac{\epsilon_{\pi^0,\gamma}}{\epsilon_{0}}\right)       &  $\epsilon^{obs}_{\pi^0,\gamma} < \epsilon^{obs}_{\pi^0,\gamma,c}$\cr
1                                                                                      &   $\epsilon^{obs}_{\pi^0,\gamma,c} < \epsilon^{obs}_{\pi^0,\gamma}$\,,\cr
}
\eary
}
and taking into account that  photons released  in the range $\epsilon_\gamma$ to $\epsilon_\gamma + d\epsilon_\gamma$ by protons in the range   $E_p$ and $E_p + dE_p$ are $f_{\pi^0}\,E_p\,(dN/dE)_p\,dE_p=\epsilon_{\pi^0,\gamma}\,(dN/d\epsilon)_{\pi^0,\gamma}\,d\epsilon_{\pi^0,\gamma}$, then photo-pion spectrum can written as
{\small
\bary
\label{pgammam}
&&\left(\epsilon^2\,\frac{dN}{d\epsilon}\right)_{\pi^0,\gamma}= A_{p,\gamma}\cr
&&\hspace{0.7cm}\times \cases{
\left(\frac{\epsilon^{obs}_{\pi^0,\gamma,c}}{\epsilon_{0}}\right)^{-1} \left(\frac{\epsilon_{\pi^0,\gamma}}{\epsilon_{0}}\right)^{-\alpha+3}          &  $ \epsilon^{obs}_{\pi^0,\gamma} < \epsilon^{obs}_{\pi^0,\gamma,c}$\cr
\left(\frac{\epsilon_{\pi^0,\gamma}}{\epsilon_{0}}\right)^{-\alpha+2}                                                                                        &   $\epsilon^{obs}_{\pi^0,\gamma,c} < \epsilon^{obs}_{\pi^0,\gamma}$\,,\cr
}
\eary
}
\noindent with the normalization energy $\epsilon_0$, proportionality constant given by
\be\label{Apg}
A_{p,\gamma}= \frac{L^{obs}_\gamma\,\epsilon^2_0\,\sigma_{peak}\,\Delta\epsilon_{peak}\left(\frac{2}{\xi_{\pi^0}}\right)^{1-\alpha}}{4\pi\,\delta_D^3\,r_d\,\epsilon^{obs}_{\gamma,b}\,\epsilon_{peak}} \,A_p\,,
\ee
\noindent and the break photon-pion energy given by 
 \be
\epsilon^{obs}_{\pi^0,\gamma,c}\simeq0.25\, \delta_D^2\,\xi_{\pi^0}\,(m_\Delta^2-m_p^2) {\epsilon^{obs}_{\gamma,b}}^{-1}\,,
\label{pgamma}
\ee
where $m_{\pi^0}$, $m_\Delta$ and $m_p$ are the pion, resonance and proton masses.   Eq. \ref{pgammam} describes the contribution of photo-pion emission to the SED.   Similarly, the proton luminosity, {\small $L_p\simeq 4\pi D^2_z F_p= 4\pi D^2_z E^2_p \frac{dN_p}{dE_p}$}, at the break photo-pion energy, $\epsilon^{obs}_{\pi^0,\gamma,c}=(\xi_{\pi^0}/2) E_p$, can be written as 
\bary\label{lum}
L_p&=& \frac{16\,\pi^2\,\delta_D^3\,r_d\,\epsilon^{obs}_{\gamma,b}\,\epsilon_{peak} D^2_z\, \left(\frac{2}{\xi_{\pi^0}}\right)^{\alpha-1}}{(\alpha-2)\,L^{obs}_\gamma\,\sigma_{peak}\,\Delta\epsilon_{peak}}\cr
&&\hspace{3.5cm}\times A_{p,\gamma}\,\left(\frac{2\epsilon^{obs}_{\pi^0,\gamma,c} }{\epsilon_0\,\xi_{\pi^0}}\right)^{2-\alpha}\,,
\eary
where  $A_{p,\gamma}$ is given by eq. \ref{Apg}.
\section{High-Energy Neutrinos}
The neutrino flux, dN$_\nu/d\en=A_{\nu} \,\en^{-\alpha_\nu}$, is correlated  with  the photon-pion spectrum by \citep[see, e.g.] [and reference therein]{2007Ap&SS.309..407H}
\be
\int \frac{dN_{\nu}}{d\en}\,\en\,d\en=\frac14\int \left(\frac{dN}{d\epsilon}\right)_{\pi^0,\gamma}\,\epsilon_{\pi^0,\gamma}\,d\epsilon_{\pi^0,\gamma}\,.
\ee
Assuming that the spectral indices for neutrino and photo-pion spectrum are similar  $\alpha\simeq \alpha_\nu$ \citep{2008PhR...458..173B}, taking into account that  each neutrino  carries 5\%  of the  proton energies ($\en=1/20\,E_p$) \citep{Halzen:2013bta} and from eq. \ref{pgammam},  we can write the relationship between HE neutrino and photon  normalization factors as 
\be\label{Anu}
A_{\nu}=\frac14A_{p,\gamma}\,\left (10\,\xi_{\pi^0}\right)^{-\alpha+2}\, {\rm TeV}^{-2},
\ee
with A$_{p,\gamma}$ given by Eq. (\ref{Apg}).    We could estimate the number of events expected per time unit (T) through
\be
N_{ev}\approx T \rho_{ice}\,N_A\, V_{eff} \int_{E_{th}}^\infty   \sigma_{\nu N}(\en)\, \frac{dN_\nu}{d\en}\,d\en,
\label{evneu1}
\ee
where E$_{th}$  is the threshold energy,   $ \sigma_{\nu N}(\en)=6.78\times 10^{-35}(\en/TeV)^{0.363}$ cm$^2=\sigma_{\nu N}(\en/TeV)^{0.363}$ is the charged current cross section \citep{1998PhRvD..58i3009G}, $\rho_{ice}\simeq$ 0.9 g cm$^{-3}$ is the density of the ice, N$_A$= 6.022$\times$ 10$^{23}$ g$^{-1}$ and V$_{eff}$ is the effective volume. If we assume that the neutrino spectrum  extends continually over the whole energy range \citep{2008PhRvD..78b3007C}, then  the expected number of neutrinos can be written as
\be\label{numneu}
N_{ev} \approx  \frac{T \rho_{ice}\,N_A\, V_{eff}}{\alpha-1.363}\,A_{\nu}\,\sigma_{\nu N}\left(\frac{E_{\nu,th}}{{\rm TeV}}\right)^{-\alpha+1.363}{\rm TeV}\,,
\ee
with A$_\nu$ given by eq. \ref{Anu}.
\section{UHE cosmic rays}
At least two events with energies larger than 60 EeV were reported and studied  by PAO inside a $3.1^{\circ}$ circle centered at Cen A \citep{2007Sci...318..938P,2008APh....29..188P}.   The study of the shower composition found that the distribution of their properties  was situated in somewhere between pure p and pure Fe at 57 EeV\citep{2008ICRC....4..335Y, 2008APh....29..188P, 2007AN....328..614U}, although  HiRes data were consistent with a dominant proton composition at these energies \citep{2007AN....328..614U}.\\
The maximum energy achieved in the comoving frame for a particle in the acceleration phase depends on the size ($r_d$)  and the strength of the magnetic field ($B'$) where it is confined,  $E'_{max}=Ze\,B'\,r_d\,$  \citep{1984ARA&A..22..425H}.  Additional limitations are mainly due to radiative losses or available time when particles diffuse through the magnetized region.\\ 
A  short distance ($\ll 1 {\rm pc}$) from the black hole (BH), the region of particle acceleration is limited by comoving dissipation radius though the variability time scale, hence the maximum energy required is \citep{2010ApJ...719.1433A,  2012PhRvD..85d3012S}
\be\label{sregion}
E_{max}=4\times 10^{19}\,{\rm eV}\, B_{0.8}\,dt^{obs}_{5}\,\Gamma_{0.85}\,,
\ee
where we have used $Q_x\equiv Q/10^x$ in c.g.s. units.  As can be seen in this small region there can not be accelerated protons up to the PAO energy range. However,  \citet{2009NJPh...11f5016D} proposed that during flaring intervals for which the apparent isotropic luminosity can reach $\approx 10^{45}$ erg s$^{-1}$ and supposing  that the  black hole  (BH) jet has the power to accelerate particles  up to ultra-high energies through Fermi processes, the maximum particle energy of accelerated UHECRs reached is
\begin{equation}
E_{max}\approx 1.0\times10^{20}\,\frac{Ze}{\phi}\frac{\sqrt{\epsilon_B\,L_{45}}}{\Gamma}\,eV,
\end{equation}
\noindent where  $\Gamma=1/\sqrt{1-\beta^2}$, $\phi\simeq 1$ is the acceleration efficiency factor and Z is the atomic number and $\epsilon_B$ comes from a equipartition magnetic field. \\
On the other hand,  \citet{2006MNRAS.368L..15H, 2007ApJ...670L..81H,  2009MNRAS.393.1041H} and \citet{2009ApJ...698.2036K} have argued that lobes are inflated by jets in the surrounding medium,  hence accelerated protons are injected into and confined within the lobes, by means of resonant Fermi-type processes, allowing that these can be re-accelerated at a distance of hundreds of kiloparsecs from the BH.  Recently, \citet{2013arXiv1312.6944F} showed that protons can be accelerated inside the lobes up to energies as high as $\sim\, 10^{20}$ eV only limited by the radius r$_d$=100 kpc (volume of $V=1.23\times 10^{71}\,cm^{3}$) and the magnetic field of  3.41 $\mu$G and 6.19 $\mu$G for the north and south lobe, respectively. Similarly,  \citet{2013A&A...558A..19W} has argued that particles in the lobes can have  acceleration stochastically by high temperatures, what would allow to reach energies as high as 10$^{20}$ eV.\\
With the mechanisms of UHECR acceleration presented above and from the correlation between the proton and $\gamma$-ray spectrum at GeV-TeV energies (eq. \ref{Apg}),  we propose that the  spectrum of accelerated protons is extended up to $\sim 10^{20}$ eV  and also that the number of these events can reach Earth. Hence, the number of UHECRs collected with PAO is calculated by means of $N_{\tiny UHECR}= ({\rm  PAO\, Expos.})\times \,N_p$, where the relative exposure from 1 January 2004 until 31 August 2007 is $\Xi\,t_{op}\, \omega(\delta_s)/\Omega_{60}=2.16/\pi\times10^4\,\rm km^2\,yr$ and N$_p=\frac{1}{(\alpha-1)}A_p \epsilon_0^\alpha\, E_p^{-\alpha+1}$ is the amount of protons at the source. Hence, from eq. (\ref{Apg}) the number of UHECRs is
{\small
\bary
N_{\tiny UHECR}\simeq 8.7\times10^4 \,{\rm km^2\,yr} \frac{\delta_D^3\,r_d\,\epsilon^{obs}_{\gamma,b}\,\epsilon_{peak}\left(\frac{2}{\xi_{\pi^0}}\right)^{\alpha-1}}{(\alpha-1)\,L^{obs}_\gamma\,\sigma_{peak}\,\Delta\epsilon_{peak}}\cr
\times \epsilon_0^{\alpha-2}\, A_{p,\gamma}\, E_{p,th}^{-\alpha+1}\,,
\label{uchers}
\eary
}
where $E_{p,th}=60$ EeV is the threshold energy of PAO.
\section{Analysis and Results}
\begin{figure}
\vspace{0.5cm}
{\centering
\resizebox*{0.8\textwidth}{0.34\textheight}
{\includegraphics{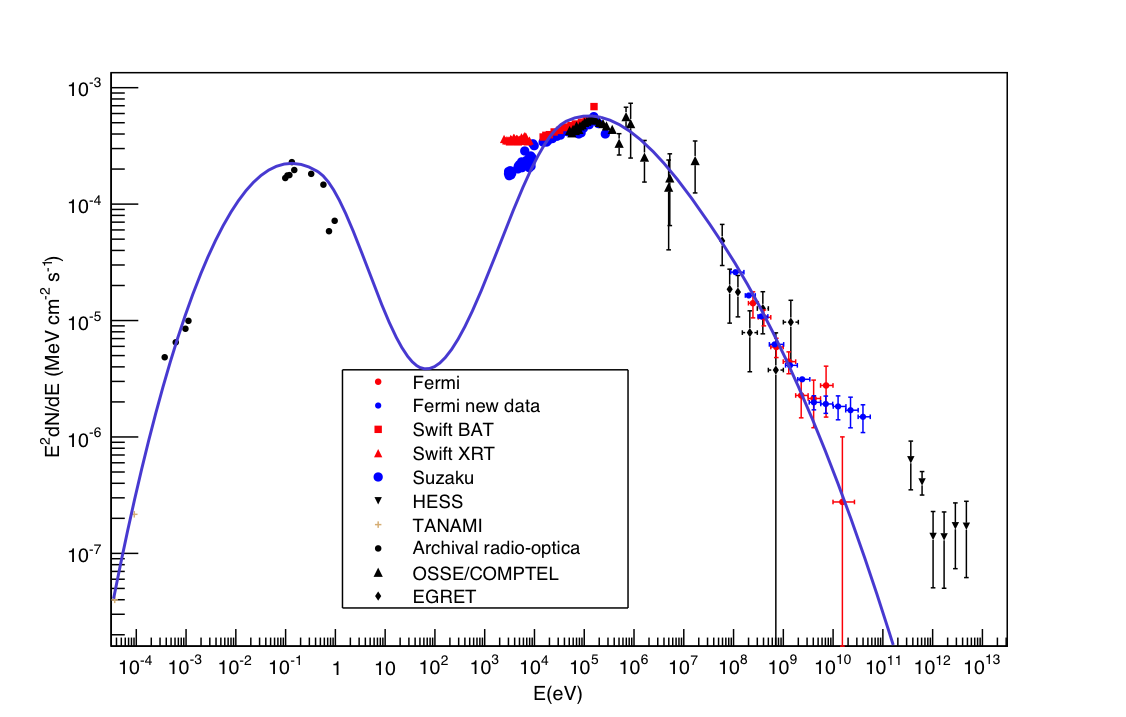}}
}
\caption{Fit of the double broad peaks of SED using a SSC leptonic model.  SED includes the Fermi data from four years of observations}
\label{Totspec}
\end{figure}

\begin{figure}
\vspace{0.5cm}
{\centering
\resizebox*{1.0\textwidth}{0.55\textheight}
{\includegraphics{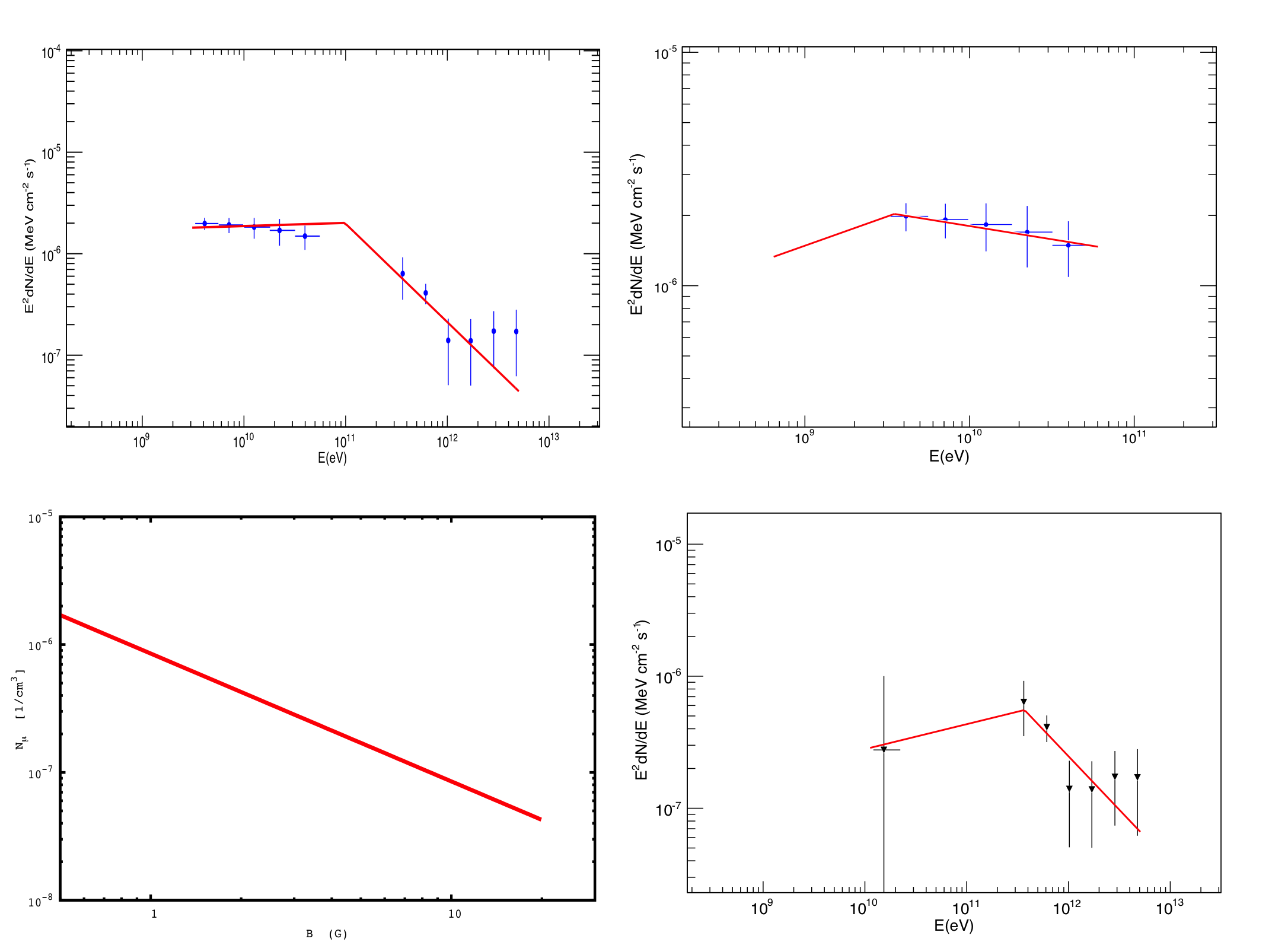}}
}
\caption{Fits of the high-energy components.  The left-hand figure above shows the fit of Fermi-LAT and H.E.S.S. data supposing that they share the same origin with the $p\gamma$ interaction.  The right-hand figure above shows  the fit of the GeV $\gamma$-ray flux with  synchrotron radiation and $p\gamma$ interaction above the break $E_b\simeq$ 4 GeV.  The left-hand figure below shows the plot of the best set of parameter $N_\mu$ muon density as a function of magnetic field with the muon synchrotron radiation model. The right-hand figure below shows the fit of the H.E.S.S. data including the last point of Fermi data}
\label{all_fit}
\end{figure}

\begin{figure}
\vspace{0.5cm}
{\centering
\resizebox*{0.65\textwidth}{0.6\textheight}
{\includegraphics{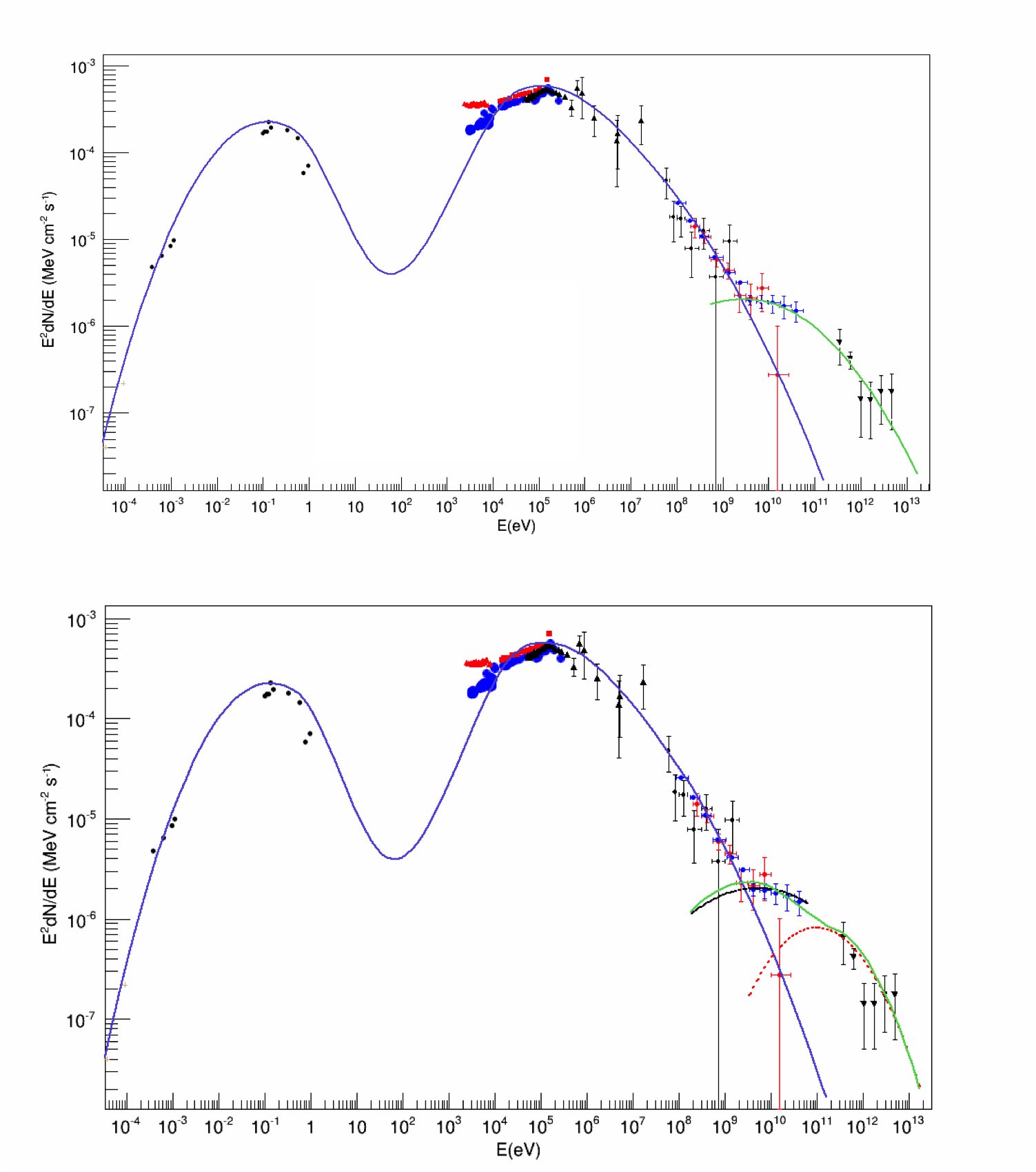}}
}
\caption{In addition to the description of the double broad peaks of SED with a SSC leptonic model (line in blue color), we have used a hadronic model based on p$\gamma$ interactions to fit the spectrum at GeV - TeV energy range. For that, we have taken into account  two seed photon populations: the intermediate (figure below)  and low state emission (figure above). In the intermediate state we have used muon synchrotron radiation (line in black color), $\pi^0$ decay products (line in red color) and the total contribution (line in green color) whereas in the low state only $\pi^0$ decay products were used (line in green color).}
\label{SED}
\end{figure}

Recently, \citet{2012ApJ...753...40F} presented a leptonic model based on SSC emission,  thereby achieving to describe successfully the two prominent humps in the spectrum of Cen A. Then, for this analysis our first approach is to use this leptonic model and include the new Fermi data  into the whole spectrum in the $\epsilon_\gamma-(\epsilon^2dN/d\epsilon)_\gamma$  representation, as shown in  figure \ref{Totspec}. In this figure can be seen that SSC model can only give account of the new Fermi-LAT data below the break energy at $\sim$4 GeV and not above it.  Hence, we will study the spectrum at higher energies than $\sim$ 4 GeV, including the  H.E.S.S. data.  We hereafter use the values  $\delta_D\simeq 1.25$,  $r_d\simeq 10^{15}\,{\rm cm}$, and  0.5 G $ \leq\, B' \leq$  10 G   which are the ones reported by  \citet{2010ApJ...719.1433A},  \citet{2012ApJ...753...40F} and  \citet{2014A&A...562A..12P}.\\ 
To interpret the $\gamma$-ray spectrum at higher energies than 4 GeV, we have introduced a hadronic model  through  the p$\gamma$ interactions and based on the fact that these interactions produce as secondary particles muons and photo-pions, we have developed a model based on synchrotron emission by radiating muons and $\pi^0$ decay products.  In the synchrotron emission, assuming that the muon spectrum is described by a broken power law, we have derived the synchrotron spectrum (eq. \ref{espsynm}). Additionally,  we have calculated the time scale characteristics and the observed photon energies for this emission process.  For our convenience, we have written the observed photon energies radiate by muons as function of those emitted by electrons (eq. \ref{ene_ph}). Thus, considering that the first broad hump was described by electron synchrotron radiation at $\epsilon^{obs}_{\gamma,c-e}\simeq 10^{-2}$ eV, then from eq. (\ref{ene_ph}),  the break energy of synchrotron radiation from short-lived muons would be $\epsilon^{obs}_{\gamma,c} \simeq 3.77$ GeV  which explains the value of break energy reported by \citet{2013ApJ...770L...6S} and  similarly, by taking into account the maximum photon energy by radiating electron $\epsilon^{obs}_{\gamma, max-e}\simeq 516.1$ MeV \citep{2012ApJ...753...40F}, then  the maximum energy radiated for this process would be $\epsilon^{obs}_{\gamma,max} \simeq 106.2$ GeV.  With the values of the break photon energies we use the method of Chi-square ($\chi^2$) minimization \citep{1997NIMPA.389...81B} to adjust Fermi data (above $\sim$ 3.77 GeV) with the observed synchrotron spectrum derived in eq. \ref{espsynm}.  In this spectrum we introduce the parameters [0]  and [1]  to obtain the best fit of  the proportionality constant ($A_{syn,\gamma-\mu}$) and power index ($\alpha$), respectively, as shown in eq. (\ref{syn_fit})
{\small
\bary
\label{syn_fit}
&&\left(\epsilon^2\,\frac{dN}{d\epsilon}\right)_{syn,\gamma}= [0]\cr
&&\hspace{0.4cm}\times  \cases {
\left(\frac{\epsilon^{obs}_{\gamma,c}}{\epsilon_0}\right)^{-1/2} \left(\frac{\epsilon_\gamma}{\epsilon_0}\right)^{-([1]-3)/2} &  $\epsilon^{obs}_\gamma < \epsilon^{obs}_{\gamma,c}$,\cr
\left(\frac{\epsilon_\gamma}{\epsilon_0}\right)^{-([1]-2)/2}           &  $\epsilon^{obs}_{\gamma,c} < \epsilon^{obs}_\gamma < \epsilon^{obs}_{\gamma,max} $\,.\cr
}
\eary
}
We show in table 1 the best set of parameters for muon synchrotron radiation and also we plot the synchrotron spectrum with the fitted parameters, as shown in fig. \ref{all_fit} (right-hand figure above).  
\begin{center}\renewcommand{\arraystretch}{1}\addtolength{\tabcolsep}{-3pt}
\begin{tabular}{ l c c c}
 \hline \hline
 \scriptsize{} &\scriptsize{Parameter} & \scriptsize{Symbol} & \scriptsize{Value}  \\
 \hline
\hline
\scriptsize{Proportionality constant} ($10^{-9}\,{\rm GeV/cm^2/s}$)  &\scriptsize{[0]}                    &   \scriptsize{$ A_{syn,\gamma-\mu} $}  &  \scriptsize{ $2.37\pm 0.611$} \\
\scriptsize{Spectral index}                                                                   &\scriptsize{[1]}                    &   \scriptsize{$\alpha$}                            & \scriptsize{2.229$\pm$ 0.0284} \\
\scriptsize{Chi-square/NDF}                                                                &                                          &    \scriptsize{$ \chi^2/{\rm NDF}$}          &  \scriptsize{ $1.867/3.0$} \\
\hline
\end{tabular}
\end{center}
\begin{center}
\scriptsize{\textbf{Table 1. The best fit of muon synchrotron radiation parameters obtained after fitting the new Fermi data.}}
\end{center}
Comparing the spectral indices; the fitted value ($\alpha$) given in table 1 and that reported ($\alpha_2$) by \citet{2013ApJ...770L...6S}, one finds that the value obtained from the fit  $\alpha'_2=(\alpha+2)/2$ = 2.114$\pm$ 0.0142  is  consistent with that reported $\alpha_2$=2.09$\pm$0.20.  Moreover,   from the best fit value of proportionally constant $A_{syn,\gamma-\mu}$, we plot the muon density as a function of magnetic field, as shown in fig. \ref{all_fit} (left-hand figure below). From this figure can be seen that for a value reported in literature of magnetic field in the range  0.5 G $\leq\, B' \leq$  10 G, the muon density range lies in $1.8\times 10^{-6}\, {\rm cm}^{-3} \leq N_\mu \leq  8.1\times 10^{-8}\, {\rm cm}^{-3}$  \citep{2010ApJ...719.1433A,2012ApJ...753...40F,  2014A&A...562A..12P}.    By considering that muons are rapidly accelerated, then the new Fermi data at the GeV energies can be described by muon synchrotron radiation.\\
From the $\pi^0$ decay model we have assumed that these neutral pions are produced in the interaction of  accelerated protons in the jet described by a simple power law with two photon populations at the emission region.  The spectrum generated by this process (eq. \ref{pgammam}) depends on the proton luminosity (through $A_p$), photon luminosity,  the comoving dissipation radius,  the break energy of observed photons, the Doppler factor and parameters of p$\gamma$ interaction.   To find the best fit of photo-pion model parameters  we use once again the method of Chi-square ($\chi^2$) minimization \citep{1997NIMPA.389...81B}, as described in eq. \ref{pgamma_fit}
{\small
\bary
\label{pgamma_fit}
&&\left(\epsilon^2\,\frac{dN}{d\epsilon}\right)_{\pi^0,\gamma}= [0]\cr
&&\hspace{0.7cm}\times \cases{
\left(\frac{\epsilon^{obs}_{\pi^0,\gamma,c}}{\epsilon_{0}}\right)^{-1} \left(\frac{\epsilon_{\pi^0,\gamma}}{\epsilon_{0}}\right)^{-[1]+3} &  $ \epsilon^{obs}_{\pi^0,\gamma}< \epsilon^{obs}_{\pi^0,\gamma,c}$\cr
\left(\frac{\epsilon_{\pi^0,\gamma}}{\epsilon_{0}}\right)^{-[1]+2}                                                                                        &   $\epsilon^{obs}_{\pi^0,\gamma,c} < \epsilon^{obs}_{\pi^0,\gamma}$\,.\cr
}
\eary
}
We will consider as target  two photon populations \citep{1998A&A...330...97S}: one population at intermediate state emission with energy peak  $\epsilon^{obs}_{\gamma,b}\sim$ 0.15 MeV and luminosity  $L^{obs}_\gamma=5\times 10^{42}$ erg/s and another at low state emission with energy peak  $\epsilon^{obs}_{\gamma,b}\sim$ 0.59 MeV and luminosity  $L^{obs}_\gamma=3\times 10^{42}$ erg/s.  As follows, we will analyze each case separately.\\
{\bf Photons at the low state emission.} Considering this photon population we calculate that the break photo-pion energy (eq. \ref{pgamma})  is $\epsilon^{obs}_{\pi^0,\gamma,c} \sim\,$ 91.3 GeV.    As shown  in fig. \ref{all_fit} (left-hand figure above) and table 2, we found  the values of A$_{p,\gamma}$ (parameter [0]) and $\alpha$ (parameter [1]) that best describe these data for $\epsilon_0$=10 GeV. 
{\small
\begin{center}\renewcommand{\arraystretch}{1}\addtolength{\tabcolsep}{-3pt}
\begin{tabular}{ l c c c}
 \hline \hline
 \scriptsize{} &\scriptsize{Parameter} & \scriptsize{Symbol} & \scriptsize{Value}  \\
 \hline
\hline
\scriptsize{Proportionality constant} ($10^{-11}\,{\rm TeV/cm^2/s}$)  &\scriptsize{[0]}                    &   \scriptsize{$ A_{p,\gamma} $}  &  \scriptsize{ $1.615\pm 0.157$} \\
\scriptsize{Spectral index}                                                                   &\scriptsize{[1]}                    &   \scriptsize{$\alpha$}                            & \scriptsize{2.970$\pm$ 0.048} \\
\scriptsize{Chi-square/NDF}                                                                &                                          &    \scriptsize{$ \chi^2/{\rm NDF}$}          &  \scriptsize{ $5.435/9.0$} \\
\hline
\end{tabular}
\end{center}
\begin{center}
\scriptsize{\textbf{Table 2.  The best fit of the $\pi^0$ spectrum  parameters obtained after fitting the new Fermi and H.E.S.S. data.}}\\
\end{center}
}
From the photo-pion spectrum (eq. \ref{pgammam}) and the value of spectral index ($\alpha$) given in table 2,  we compare both power laws, below and above of  $\epsilon^{obs}_{\pi^0,\gamma,c} \sim\,$ 91.3 GeV with new Fermi and H.E.S.S. data, respectively.  Comparing the power law below $\epsilon^{obs}_{\pi^0,\gamma,c}$, we can see that the obtained spectral index is $\alpha'_2=\alpha-1$ = 1.970$\pm$ 0.048 which is consistent to that reported ($\alpha_2$=2.09$\pm$0.20) by \citet{2013ApJ...770L...6S} and when we compare the power law above $\epsilon^{obs}_{\pi^0,\gamma,c}$, then the obtained spectral index is $\alpha'_2=\alpha$ = 2.970$\pm$ 0.048 which is also in agreement with the reported value  2.7$\pm\, 0.5_{stat} \pm 0.2_{sys}$ by \citet{2009ApJ...695L..40A}.
Once again from the fitted values given in table 2 and eqs. (\ref{numneu}) and (\ref{uchers}), we calculate the number of neutrinos and UHECRs expected on IceCube and PAO, respectively.  Assuming a threshold energy of  E$_{th,\nu}$ = 30 TeV and  two years of observation \citep{2013arXiv1311.5238I}, we obtained that 2.37 $\times 10^{-2}$ events are expected in IceCube, and considering a simple power law extended as high as PAO energy range, we  found that 3.93 events are expected on PAO.  In addition to the analysis performed,  a proton luminosity of  2.2 $\times 10^{44}$ erg/s is required.\\
{\bf Photons at intermediate state emission.} Assuming this photon population, a similar analysis to the previous case will be performed. As has been pointed out, Cen A exhibits two  prominent humps, the first one is related with photons at low energies 10$^{-2}$ eV  and the second one to those at high energies 150 keV.    In this case we assume that accelerated protons interact with photons of around $\sim$ 150 keV which have a photon density (eq. \ref{den}) of 5.52$\times$ 10$^7$ cm$^{-3}$  and an optical depth (eq. \ref{opt}) of  2.76$\times$ 10$^{-3}$.  Then, from eq. (\ref{pgamma}),  the break photo-pion energy is $\epsilon^{obs}_{\pi^0,\gamma,c} \sim$ 359.11 GeV.   As shown  in fig. \ref{all_fit} (right-hand figure below) and table 3, we found  the values of A$_{p,\gamma}$ (parameter [0]) and $\alpha$ (parameter [1]) that best describes these data for $\epsilon_0$=1 TeV.
\begin{center}\renewcommand{\arraystretch}{1}\addtolength{\tabcolsep}{-3pt}\label{fit_pion1}
\begin{tabular}{ l c c c}
  \hline \hline
 \scriptsize{} &\scriptsize{Parameter} & \scriptsize{Symbol} & \scriptsize{Value}  \\
 \hline
\hline
\scriptsize{Proportionality constant} ($10^{-13}\,{\rm TeV/cm^2/s}$)  &\scriptsize{[0]}                    &   \scriptsize{$ A_{p,\gamma} $}  &  \scriptsize{ $2.478\pm 0.492$} \\
\scriptsize{Spectral index}                                                                   &\scriptsize{[1]}                    &   \scriptsize{$\alpha$}                            & \scriptsize{2.811$\pm$ 0.378} \\
\scriptsize{Chi-square/NDF}                                                                &                                          &    \scriptsize{$ \chi^2/{\rm NDF}$}          &  \scriptsize{ $3.279/5.0$} \\
\hline
\end{tabular}
\end{center}
\begin{center}
\scriptsize{\textbf{Table 3.  The best fit of the set of p$\gamma$ interaction parameters obtained  after fitting the TeV spectrum.}}\\
\end{center}
As shown in table 3,  the fitted value of the spectral index ($\alpha=2.811\pm 0.378$)  and the reported one by H.E.S.S. \citep{ 2009ApJ...695L..40A}  2.7$\pm\, 0.5_{stat} \pm 0.2_{sys}$ are in agreement.   Also from this plot (right-hand figure below) can be seen that  the data value  at energy of $15.36$ GeV fits nicely to the photo-pion emission.  In a like manner,  from the values of table 3 and eqs. (\ref{numneu}) and (\ref{uchers}),  we estimate  that the number of neutrinos  and UHECRs expected are 0.82$\times 10^{-3}$ and 10.44 events on IceCube and PAO, respectively. Again we have assumed that the proton power law is extended  up to PAO energy range. For this case, a proton luminosity of 2.03 $\times 10^{43}$ erg/s is required.\\
In addition to the SSC leptonic model \citep{2012ApJ...753...40F} and taking into account the p$\gamma$ interactions with both photon populations as targets, we plot the whole SED of Cen A using our leptonic and hadronic model, as shown in fig. \ref{SED}. In the figure above we use seed photons at low state emission and in figure below the intermediate state emission was taken into account.  From both figures we can see that unlike emission of  $\pi^0$ decay products which is asked for when regarding both photon populations,  a muon synchrotron radiation is required when we consider photons at intermediate state emission but not when considering photons at low state emission in order to explain successfully the spectrum at GeV - TeV energy range.\\
Finally, analyzing both contributions of $\pi^0$ decay products at the same time (intermediate and slow state emission as seed photons) and regarding $\epsilon^{obs}_{\pi^0,\gamma,hl}<\epsilon^{obs}_{\pi^0,\gamma,cl}$ with $\epsilon^{obs}_{\pi^0,\gamma,cl}$= 91.3 GeV and $\epsilon^{obs}_{\pi^0,\gamma,ch}$= 359.11 GeV, the photo-pion spectrum (eq. \ref{pgammam}) would have a small change; a power law would be added to the spectrum $\left(\epsilon^2\, dN/d\epsilon\right)_{\pi^0,\gamma}\simeq\, A_{p,\gamma} \left[ 1+\left(\epsilon_{\pi^0,\gamma}/ \epsilon^{obs}_{\pi^0,\gamma,ch} \right) \right]\,\left(\epsilon_{\pi^0,\gamma}/\epsilon_0 \right)^{-\alpha+2}$ for $\epsilon^{obs}_{\pi^0,\gamma,cl} < \epsilon_{\pi^0,\gamma}<\epsilon^{obs}_{\pi^0,\gamma,ch}$  whereas the first and the last power laws can be approximated to same ones (eq. \ref{pgammam});   $\simeq  \left(\epsilon^{obs}_{\pi^0,\gamma,cl}/\epsilon_{0}\right)^{-1} \left(\epsilon_{\pi^0,\gamma}/\epsilon_{0}\right)^{-\alpha+3}$ for $ \epsilon_{\pi^0,\gamma} < \epsilon^{obs}_{\pi^0,\gamma,c}$  and   $\left(\epsilon_{\pi^0,\gamma}/\epsilon_{0}\right)^{-\alpha+2}$ for  $\epsilon^{obs}_{\pi^0,\gamma,ch} < \epsilon_{\pi^0,\gamma}$, where we have assumed that the contribution of both interactions to the spectrum are similar.  As a result, the fitted values are not much pretty different from those given in table 2,  then we would expect small changes in the values of the proton proportionality constant and spectral index, and thus in the numbers of UHECRs and neutrinos in comparison with those values obtained when considering the low state emission as photon population in the p$\gamma$ interactions.
\section{Summary and conclusions}
We have proposed a leptonic and hadronic model to explain the $\gamma$-ray spectrum at GeV - TeV energy range. In the leptonic model, we have applied the SSC emission described by \citet{2012ApJ...753...40F} and have showed that  SSC emission only gives account of photons with energies of less than $\approx$ 4 GeV. To explain the spectrum at higher energies than 4 GeV, we have introduced  the hadronic model,  assuming that accelerated protons in the jet interact with two photon population; at intermediate and low state emission.  Based on these p$\gamma$ interactions, we have developed synchrotron radiation by muons and $\pi^0$ decay products.  Considering the photon population at low state emission, we found that both Fermi and H.E.S.S. data are well described by means of photo-pion spectrum (eq. \ref{pgammam}) with break energy of 91.3 GeV and proton luminosity of  2.2$\times 10^{44}$ erg/s.   In this case, the contribution of  synchrotron radiation to the $\gamma$-ray spectrum is not required, so both new Fermi and H.E.S.S. data are described by $\pi^0$ decay products.   Additionally, we have estimated that the numbers of UHECRs and neutrinos  are  3.93 and 2.37 $\times 10^{-2}$, respectively.  On the other hand,  assuming the photon population with intermediate state, H.E.S.S. and Fermi data are described separately;  new Fermi data with muon synchrotron radiation and H.E.S.S. data with $\pi^0$ decay products. From this emission, we found the values of  break energy of 359.11 GeV and  proton luminosity of 2.03$\times 10^{43}$ erg/s. Additionally, extending the proton and neutrino spectrum  again up to UHE, we estimate  that the number of UHECRs is 10.44 and that of neutrinos 0.82 $\times 10^{-3}$.  It has to be added that although muons radiate at GeV energy range by synchrotron emission,  if muons do not radiate rapidly, this process would not be effective, then $\pi^0$ decay products originated from low emission state would describe the whole spectrum at GeV - TeV energy range.\\
On the other hand, we have showed that extrapolating the proton spectrum by a simple power law up to 10$^{20}$ eV, the number of UHECRs expected is closer when considering the low (3.93) than intermediate (10.44) state emission.  Although UHECRs can hardly be accelerated up to the PAO energy range at the emission region (E$_{max}$= 40  EeV) \citep{2010ApJ...719.1433A}, they could be accelerated during the flaring intervals and/or in the giant lobes.  It is very interesting the idea that UHECRs could be accelerated partially in the jet at energies ($<40\times 10^{19}$ eV) and partially in the Lobes at ($E>40\times 10^{19}$ eV).\\
In summary,  we have showed that hybrid leptonic SSC and hadronic processes are required to explain the $\gamma$-fluxes at GeV- TeV energy range. We have successfully described the spectral indices, break energies and fluxes, and also the expected number of  UHECRs and neutrinos \citep{2009ApJ...695L..40A, 2013ApJ...770L...6S,2008APh....29..188P, 2013arXiv1311.5238I}. 
\section*{Acknowledgements}
We thank to Bing Zhang, Francis Halzen and William Lee for useful discussions.  This work was supported by Luc Binette scholarship and the projects IG100414 and Conacyt 101958.
%

%
\clearpage
\clearpage
\clearpage
\end{document}